# Implementation of Artificial Neural Networks for the Nepta-Uranian Interplanetary (NUIP) Mission

*Saurabh Gore* [(1)], *Manuel Ntumba* [(2)]

[(1)(2)] *Division of Space Applications, Tod'Aérs - Transparent Organization for Aeronautics and Space Research, Togo.*

[(1)] *Moscow Aviation Institute, Moscow, Russia.*

[(2)] *Space Generation Advisory Council, Vienna, Austria.*

manuel.ntumba@spacegeneration.org

**ABSTRACT**

A celestial alignment between Neptune, Uranus, and Jupiter will occur in the early 2030s, allowing a slingshot around Jupiter to gain enough momentum to achieve planetary flyover capability around the two ice giants. The launch of the uranian probe for the departure windows of the NUIP Mission is between January 2030 and January 2035, and the duration of the mission is between 6 and 10 years, and the launch of the Nepta probe for the departure windows of the NUIP Mission is between February 2031 and April 2032 and the duration of the mission is between 7 and 10 years. To get the most out of alignment, deep learning methods are expected to play a critical role in autonomous and intelligent spatial guidance problems. This would reduce travel time, hence mission time, and allow the spacecraft to perform well for the life of its sophisticated instruments and power systems up to 15 years. This article proposes a design of deep neural networks, namely convolutional neural networks (CNN) and recurrent neural networks (RNN), capable of predicting optimal control actions and image classification during the mission. Nepta-Uranian interplanetary mission, using only raw images taken by optimal onboard cameras. It also describes the unique requirements and constraints of the NUIP mission, which led to the design of the communications system for the Nepta-Uranan spacecraft. The proposed mission is expected to collect telemetry data on Uranus and Neptune while performing the flyovers and transmit the obtained data to Earth for further analysis. The advanced range of spectrometers and particle detectors available would allow better quantification of ice giants' properties.

## 1. INTRODUCTION

Voyager-2 was the fourth of five spacecraft to achieve solar escape speed, which will allow it to leave the solar system. NASA launched Voyager-2 in 1977 to study the outer planets. It was launched 16 days before Voyager 1, part of the Voyager on a Path program that allowed new encounters with Uranus and Neptune. [1] Voyager 2 has been operating for 43 years, and its mission is now in its extended mission to study interstellar space. He stays in contact via NASA's distant space network. [2] Contact was reestablished in 2020 when a series of instructions were transmitted, then executed and relayed with a successful communication message. [3] Full communications with the probe were reestablished after a major antenna upgrade, which took a year, from the communications link, solely responsible for communications with the probe, to Canberra, Australia. [4] Voyager's missions have dramatically improved our understanding of Galilee's moons and discovered the rings of Jupiter. They also confirmed that the big red dot was high pressure. Images showed that the red dot had changed color from the Pioneer missions, from orange to dark brown. As the spacecraft passed behind the planet Jupiter, it observed lightning in the nighttime atmosphere. [15] [17] The next mission to meet Jupiter was Ulysses' solar probe. He performed an overflight maneuver to reach a polar orbit around the Sun. During this passage, the spacecraft studied Jupiter's magnetosphere. Ulysses does not have a camera, so no images were taken. [18] In 2000, the Cassini probe flew over Jupiter en route to Saturn and provided higher resolution images. [19] The New Horizons probe flew through Jupiter in 2007 for gravity assistance en route to Pluto.[20]

## 2. MISSION OVERVIEW

For the NUIP mission, the Hohmann transfer is used to move the Nepta-Uranian spacecraft from Earth orbit to Jupiter's orbit. The Hohmann transfer orbit moves between two circular orbits in the same plane and travels exactly 180 ° around the primary. An orbit that travels less than 180 ° around the primary is called Hohmann [13]. The alignment of the planets in their orbits is crucial for the Hohman transfer. The alignment requirement gave birth to the concept of launch windows. At launch, Nepta-Uranian has muzzle velocity and kinetic energy associated with its orbit around the Earth. Therefore, relatively small thrusts at either end of the path are needed to organize the transfer. The use of CNN to optimize the classification of images to avoid rapid speed when passing through Jupiter. Dual-space cameras are taking quick photos using CNN features. Jupiter is a gas giant whose mass is one-thousandth of that of the Sun. [14] Several spacecraft performed planetary flyover maneuvers that brought them within the observation range of Jupiter. Images of the atmosphere of Jupiter and several of its moons were obtained. They found that the radiation fields near the planet were much stronger than expected, but both spacecraft managed to survive in this environment. Sending state-of-the-art instruments and sensors to two probes would greatly increase our knowledge of the two ice giants. Ice giants represent one of two different categories of giant planets found in the solar system. The other group is the best-known gas giants, consisting of over 90% hydrogen and helium (by mass). Ice giants are mostly made of heavier elements. Their hydrogen is believed to extend to their small rocky cores, where molecular hydrogen ions transform into metallic hydrogen under hundreds of gigapascals of extreme pressures (GPa). [5] While ice giants also have hydrogen envelopes, these are much smaller. They represent less than 20% of their mass. Oxygen, carbon, nitrogen, and sulfur are more abundant in the universe. The hydrogen of the ice giants never reaches the depths necessary for the pressure to create metallic hydrogen. [5] Nevertheless, these envelopes limit the observation of the ice giants' interiors, and therefore the information on their composition and evolution. [5] Uranus and Neptune are called giant ice planets. However, it is believed that there is an ocean of supercritical water beneath their clouds, which makes up about two-thirds of their total mass. [6] [7] weather conditions. It forms and dissipates every few years, unlike Jupiter's large red spot of similar size, which has persisted for centuries. Neptune emits the internal heat per unit of sunlight absorbed. A ratio of about Saturn, the second-highest emitter, only has a ratio of about 1.8. The biggest feature visible on Neptune is the recurring large dark spot. Uranus emits the least heat, a tenth less than Neptune. This makes its seasonal patterns very different from those of any other planet in the solar system.[5] Understanding these characteristics will help elucidate how the atmospheres of giant planets work in general. [5] Therefore, such information could help scientists better predict the atmospheric structure and behavior of giant exoplanets found very close to their host stars, Pegasian planets, and exoplanets whose masses and rays fall from giant planets and terrestrials found in the solar system. [5] Due to their large dimensions and low thermal conductivity, internal planetary pressures range up to several hundred GPa and temperatures up to several thousand Kelvin (K). [8]

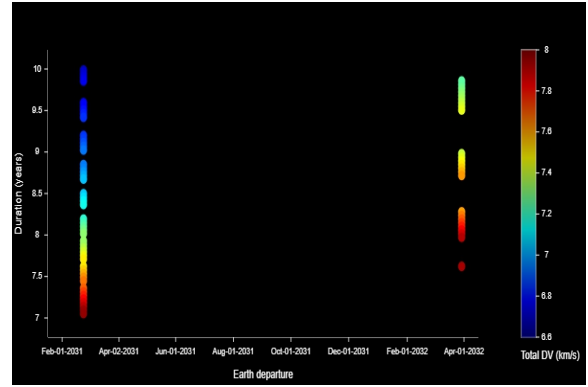

**Figure 1: Launch windows of Nepta spacecraft.** Nepta Mission's launch for the NUIP Mission is described in Figure 1. The mission Earth departure windows are between February 2031 and April 2032. And the duration of the mission is between 7 and 10 years

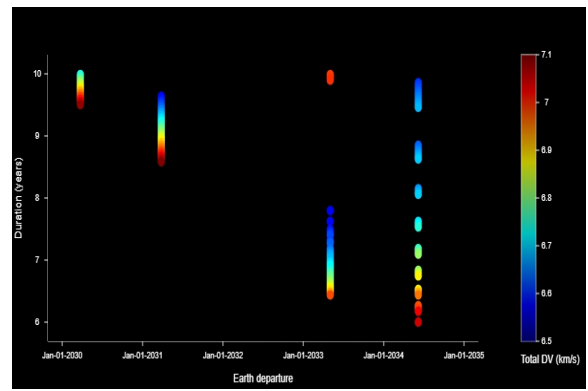

**Figure 2: Launch windows of Uranian spacecraft.** Uranian Mission's launch for the NUIP Mission is described in Figure 2. The mission Earth departure windows are between January 2030 and January 2035. And the duration of the mission is between 6 and 10 years.

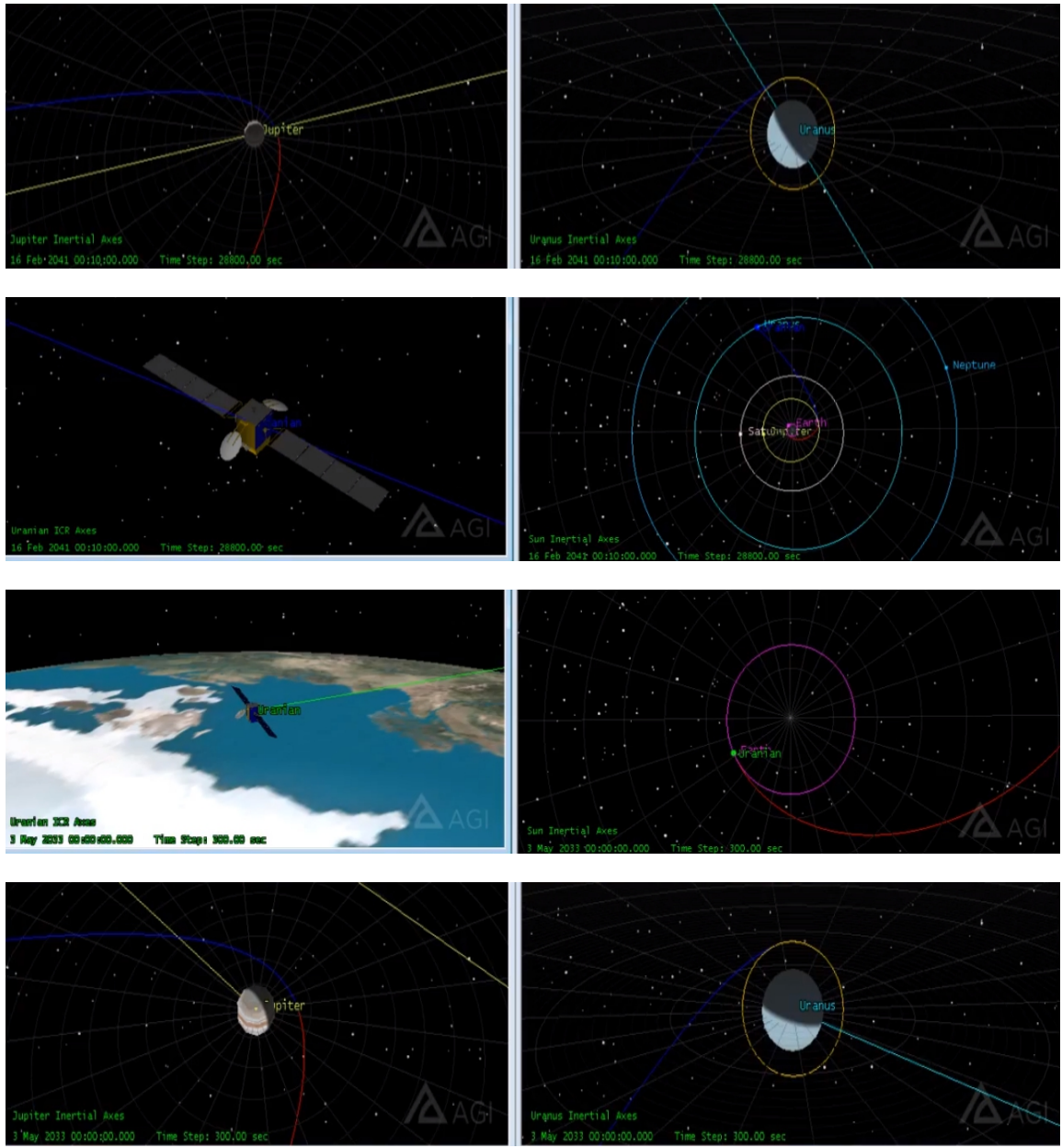

**Figure 3: Digital Mission Analysis of Uranian spacecraft from Jupiter to Uranus.** It was proven in 2012 that the compressibility of water used in models of ice giants could be reduced by a third. [9] This value is important for the modeling of ice giants and has a ripple effect on their understanding. [9] The intensities of Uranus and Neptune's magnetic fields are unusually shifted and are intermediate between those of gas giants and those of terrestrial planets, respectively 50 and 25 times that of the Earth. The intensities of the equatorial magnetic field of Uranus and Neptune are 75% and 45% of 0.305 gauss of Earth, respectively. Their magnetic fields are believed to originate from a mantle of fluid ice with ionized convection. [10] The spacecraft's cameras measured plasma output from volcanoes on Io and studied the four Galilean moons in detail, as well as long-range observations of the outer moons Himalia and Elara.[21]

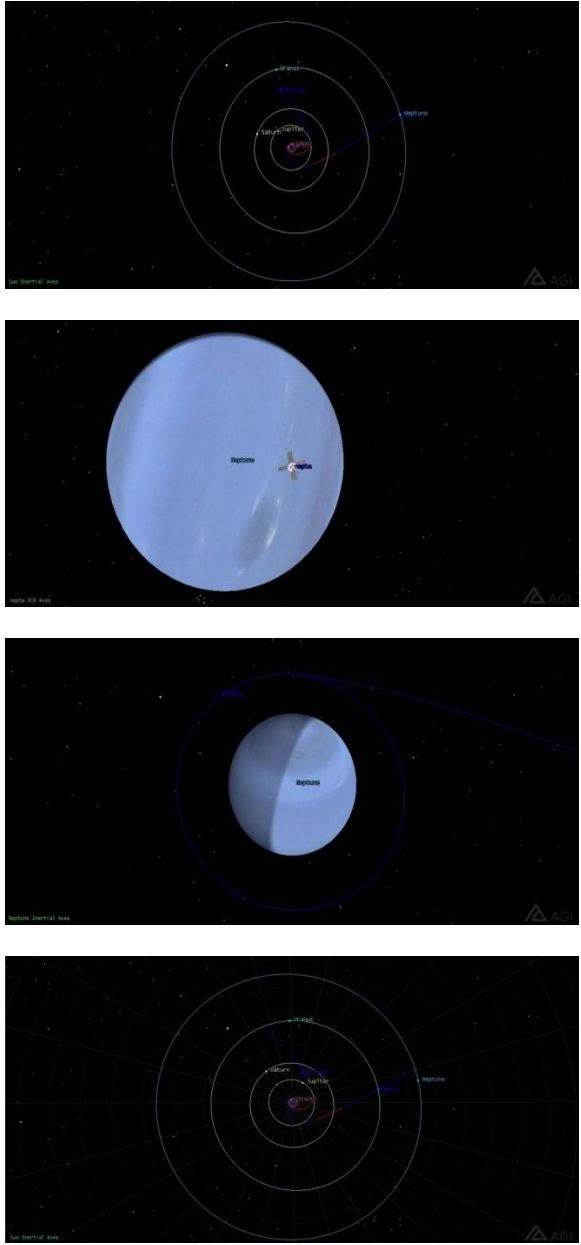

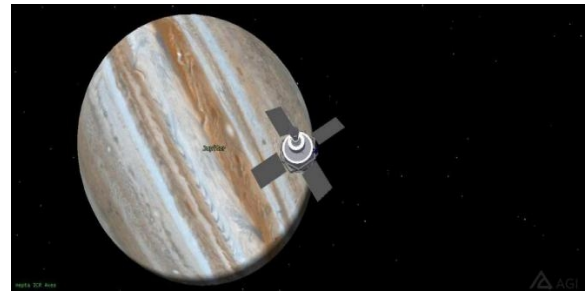

**Figure 4: Digital Mission Analysis of Nepta spacecraft from Jupiter to Neptune.** The outer gas layers of ice giants have several similarities to gas giants. These include high-speed and long-lasting equatorial winds, polar eddies, complex chemical processes driven by ultraviolet radiation from above, large-scale circulation patterns, and lower atmosphere. [5] The study of the atmospheric configuration of ice giants also provides their compositions with different chemical processes. They receive much less sunlight in their distant orbits than any other planets in the solar system, increasing the relevance of internal heating to

**Figure 5: NUIP Mission during Jupiter flyby.** To complete the mission within a reasonable time, it is important to use Jupiter's gravity to our advantage. The mission is relatively achievable as it will significantly reduce fuel requirements thanks to Jupiter's efficient slingshot window. The trajectories are optimized for better optical communication with the ground station. [15] [16] and the use of RNNs to optimize trajectory parameters. The Nepta-Uranian needs a certain speed to orbit Jupiter, which will be less than the speed needed to continue to orbit the Sun in transfer orbit, let alone attempt to orbit the Sun in orbit. Therefore, the spacecraft will have to decelerate for Jupiter's gravity to capture it. This capture etching should be done optimally at low altitudes to get the most out of the Oberth effect.

## 3. METHODOLOGY

Long-term memory (LSTM) is an artificial neural network architecture [39] used in deep learning. LSTM has feedback connections and can process single data points such as images and entire data sequences such as video. LSTM applies to tasks such as non-segmented connected handwriting recognition, [40] speech recognition [41] [42], and detection of anomalies in network traffic or IDS detection systems. A common LSTM unit consists of a cell, an entry door, an exit door, and a forgetting door. The model remembers the values over arbitrary time intervals and the three gates regulate the flow of information in and out of the cell. LSTM networks are well suited for classifying, processing, and making predictions based on time series data, as there can be lags of unknown duration between important events in a time series. There is also a possibility of coplanar mono-injection. The calculation of the vector of plane B from a position vector-centered on the planet r and a speed vector v at the closest approach to the planet flown over.[43]

$$\mathbf{v}_1 = \frac{1}{2} v_\infty \left[ (D+1)\hat{\mathbf{i}}_\infty + (D-1)\hat{\mathbf{i}}_r \right]$$

where

$$D = \sqrt{1 + \frac{4\mu}{rv_\infty^2 (1 + \hat{\mathbf{i}}_\infty \cdot \hat{\mathbf{i}}_r)}}$$

and

$$\hat{\mathbf{i}}_r = \begin{bmatrix} \cos\Omega\cos\theta - \sin\Omega\sin\theta\cos i \\ \sin\Omega\cos\theta + \cos\Omega\sin\theta\cos i \\ \sin\theta\sin i \end{bmatrix}$$

$$\hat{\mathbf{i}}_\theta = \begin{bmatrix} -\cos\Omega\sin\theta - \sin\Omega\cos\theta\cos i \\ -\sin\Omega\sin\theta + \cos\Omega\cos\theta\cos i \\ \cos\theta\sin i \end{bmatrix}$$

The speed vector of the spacecraft in the initial circular orbit is given by $v_o$. The velocity vector at any geocentric position vector r necessary to realize a launching hyperbola [44] [45] The specific launch, overflight and destination planets, and launch predictions, overflight parameters for the overflight altitude. The Delta-V required at launch and arrival is simply the difference between the speed on the transfer path determined by the Lambert problem's solution and the two planets' heliocentric speeds.

The two vector equations give the required maneuvers. The calculation steps to create an initial estimate of the state of the spacecraft, and the launch and delta-v arrival characteristics to calculate the starting planet state vector at the initial estimated departure date, calculate the state vector of the hovered planet at the initial assumption of the hover date, calculate the state vector of the arrival planet at the initial estimate of the arrival date, and solve the problem of Lambert.

semimajor axis

$$a = \frac{r}{\left(2 - \frac{rv^2}{\mu}\right)}$$

orbital eccentricity

$$e = \sqrt{1 - p/a}$$

true anomaly

$$\cos\theta = \frac{p-r}{er} \qquad \sin\theta = \frac{\dot{r}h}{e\mu}$$

B-plane magnitude

$$B = \sqrt{p|a|}$$

fundamental vectors

$$\hat{\mathbf{z}} = \frac{r\mathbf{v} - \dot{r}\mathbf{r}}{h}$$

$$\hat{\mathbf{p}} = \cos\theta\,\hat{\mathbf{r}} - \sin\theta\,\hat{\mathbf{z}}$$

$$\hat{\mathbf{q}} = \sin\theta\,\hat{\mathbf{r}} + \cos\theta\,\hat{\mathbf{z}}$$

$$\hat{\mathbf{q}} = \sin\theta\,\hat{\mathbf{r}} + \cos\theta\,\hat{\mathbf{z}}$$

S vector

$$\mathbf{S} = -\frac{a}{\sqrt{a^2 + b^2}}\hat{\mathbf{p}} + \frac{b}{\sqrt{a^2 + b^2}}\hat{\mathbf{q}}$$

B vector

$$\mathbf{B} = \frac{b^2}{\sqrt{a^2 + b^2}}\hat{\mathbf{p}} + \frac{ab}{\sqrt{a^2 + b^2}}\hat{\mathbf{q}}$$

T vector

$$\mathbf{T} = \frac{\left(S_y^2, -S_x^2, 0\right)^T}{\sqrt{S_x^2 + S_y^2}}$$

R vector

$$\mathbf{R} = \mathbf{S} \times \mathbf{T} = \left(-S_z T_y,\ S_z T_x,\ S_x T_y - S_y T_x\right)^T$$

angular momentum vector
$$h = r \times v$$

radius rate
$$\dot{r} = \frac{r \cdot v}{|r|}$$

semiparameter
$$p = \frac{h^2}{\mu}$$

The DNNs algorithms implemented in the models assume that the injection occurs impulsively at the starting hyperbola's perigee. These DNNs are valid for geocentric orbit inclinations, which satisfy the geometric constraint on the orbit's orbital inclination. DNNs are also implemented to design and optimize gravity-assisted conical interplanetary trajectories to design and optimize conflict-corrected interplanetary trajectories that include a single gravitational assist maneuver. For the NUIP mission, the implemented DNN algorithms assume an impulsive injection maneuver that only involves an instantaneous change of speed. The orbital inclination of the Earth's fleet's initial circular orbit and the outgoing or launching hyperbola's declination. There are two planetary opportunities to establish a starting hyperbola to satisfy the outgoing path's energy and orientation, and an injection opportunity will occur when the spacecraft ascends and the other spacecraft descends alongside it.

## 4. RESULTS AND DISCUSSION

The implementation of DNNs to improve scientific investigations, reduce cost, and reduce mission time, to shorten travel times to Uranus and Neptune, the determining factor is the alignment. As the journey times were shortened, the approach speed necessarily increased the mass of a propulsion system using conventional space storable thrusters, increasing exponentially with approach speed. Methods of increasing the propulsion system's specific impulse, which is the denominator of this exponential factor through hydrogen propulsion and electric radioisotope propulsion. Design of integrated deep neural networks capable of predicting the magnitude and direction of optimum thrust for fuel directly from a sequence of images taken by the on-board camera system. A combination of deep CNN and RNN-LSTM can be trained to determine the optimal fuel guidance. Such an approach requires calculating off-line, fuel-efficient open-loop trajectories and simulating optical images taken from an on-board camera along with the NUIP Mission. Thus, a training set can be generated, and deep networks can be formed to improve the optimal guidance and determine the functional relationship between the sequence of images and the optimal thrust. In this case, a CNN was trained and tested to predict the level of accuracy. The evolution of classification and regression losses during the deep RNN formation phase can be overcome by using DNNs to predict both thrust magnitude and thrust direction based on a sequence of images and atmospheric compositions.

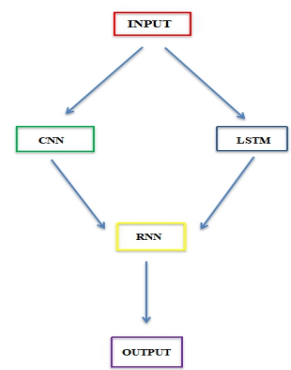

**Figure 6: Proposed Architecture of the Artificial Neural Networks.** LSTMs were developed to solve the leakage gradient problem that can be encountered during the formation of traditional RNNs. The relative gap length insensitivity is advantageous for LSTM over RNNs, hidden Markov models, and other sequence learning methods in many applications. In RNN, the connections between nodes form an oriented graph along a temporal sequence. This allows it to exhibit a temporal dynamic behavior. Derived from direct-acting neural networks, RNNs can use their internal state memory to process input sequences of varying lengths. [22] [23] [24] This makes them applicable to tasks such as non-segmented connected handwriting recognition [25] or speech recognition. [26] [27] In deep learning, a convolutional neural network (CNN) is a deep neural network class most commonly applied to visual imagery analysis. [28] Also called Space Invariant Artificial Neural Networks (SIANN), based

on the weight-shared architecture of convolution nuclei that sweep through the body's hidden layers and features. invariance of translation. [29] [30] They have image and video recognition applications, recommendation systems, [31] image classification, image segmentation, medical image analysis, and patient treatment. natural language, [32] brain-computer interfaces, [33] and financial time series. [34] CNNs are regularized versions of multilayered perceptrons because every neuron in one layer is connected to all neurons in the next layer. The connectivity of these networks makes them prone to data overfitting. Typical means of regularization include varying the weights as the loss function is minimized while reducing connectivity at random. CNNs assemble models of increasing complexity using smaller, simpler models embossed in filters. Therefore, in terms of connectivity and complexity, CNNs are at the very bottom. Convolutional networks have been inspired by biological processes [35] [36] [37] [38] in that the pattern of connectivity between neurons resembles the natural visual cortex.

The data visualizations for the NUIP Mission are based on the astrodynamic model of NUIP Mission for both Nepta spacecraft and Uranian spacecraft. Medium Fidelity Parameters are mentioned in Table 1 and Table 4. Targeter Control variables are mentioned in Table 2 and Table 5. The orbital parameters are mentioned in Table 3 and Table 6.

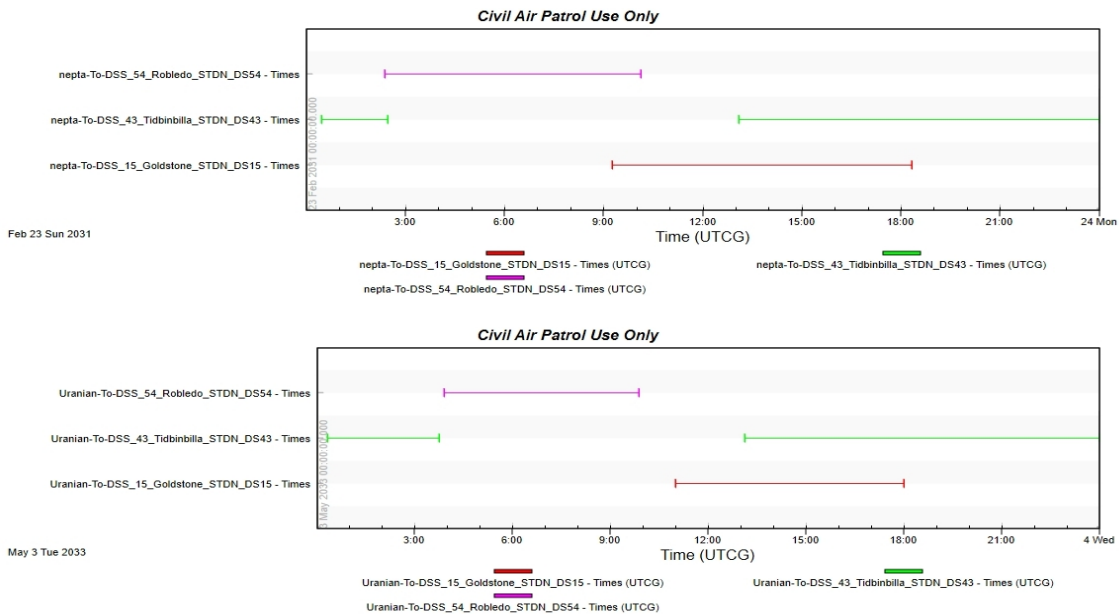

Figure 7: Cumulative Time intervals of Nepta-Uranian (NUIP) Mission.

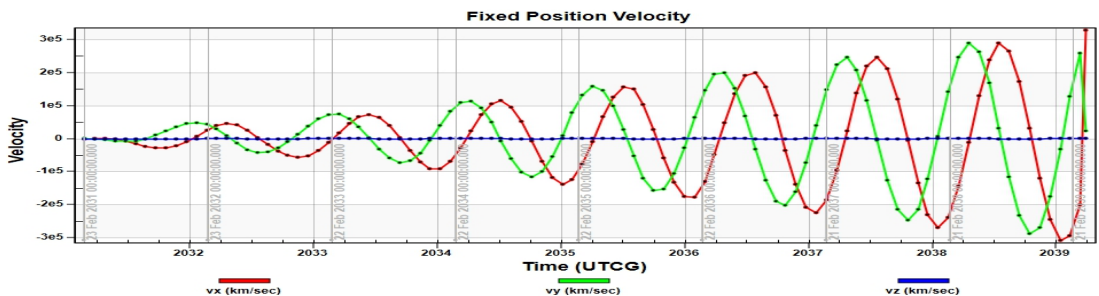

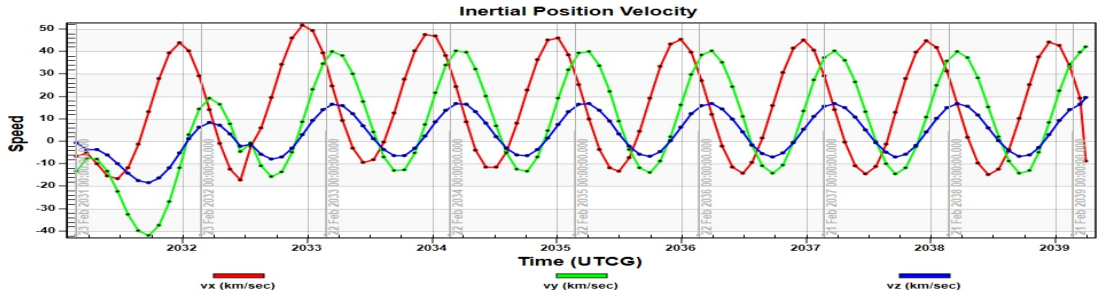
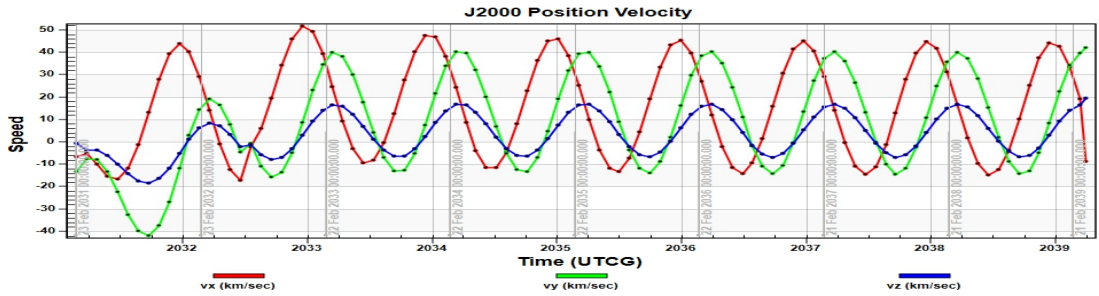
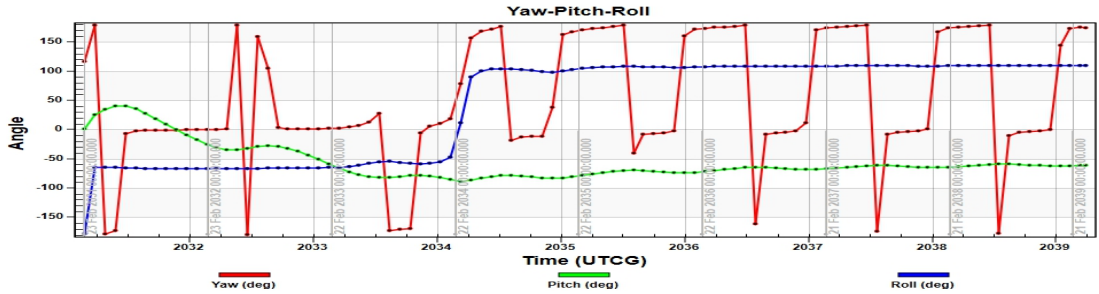
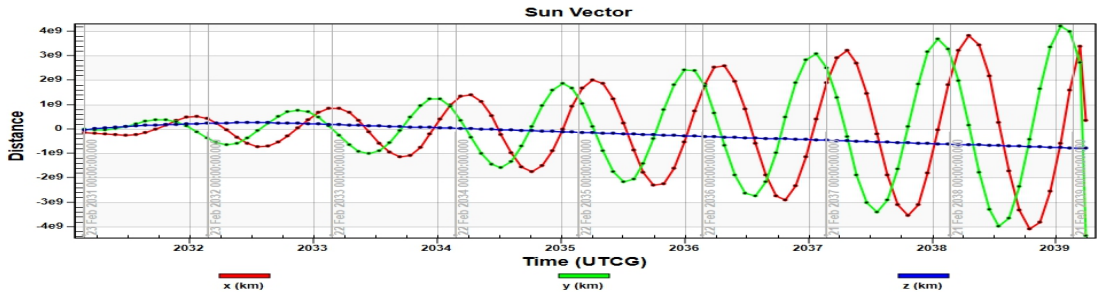
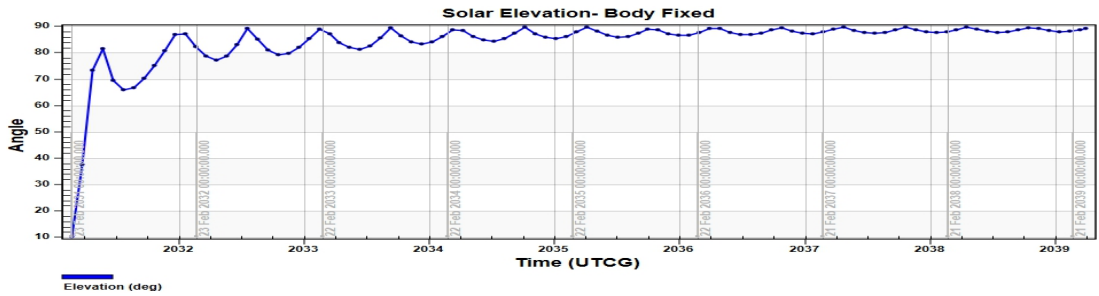

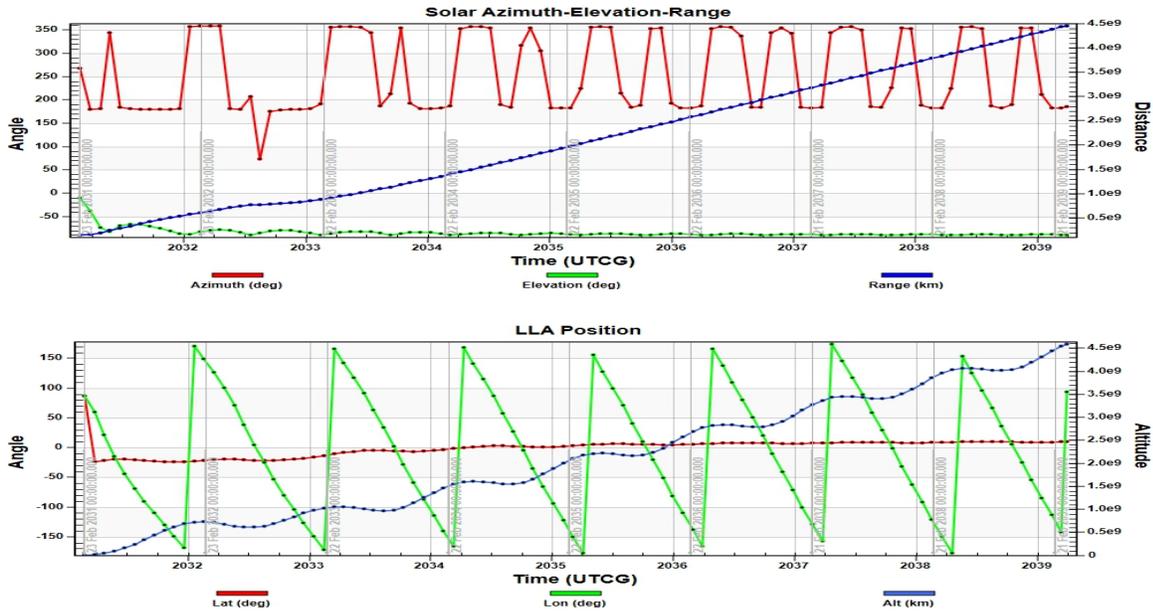

Figure 8: Uranian's Targeter Control Variables.

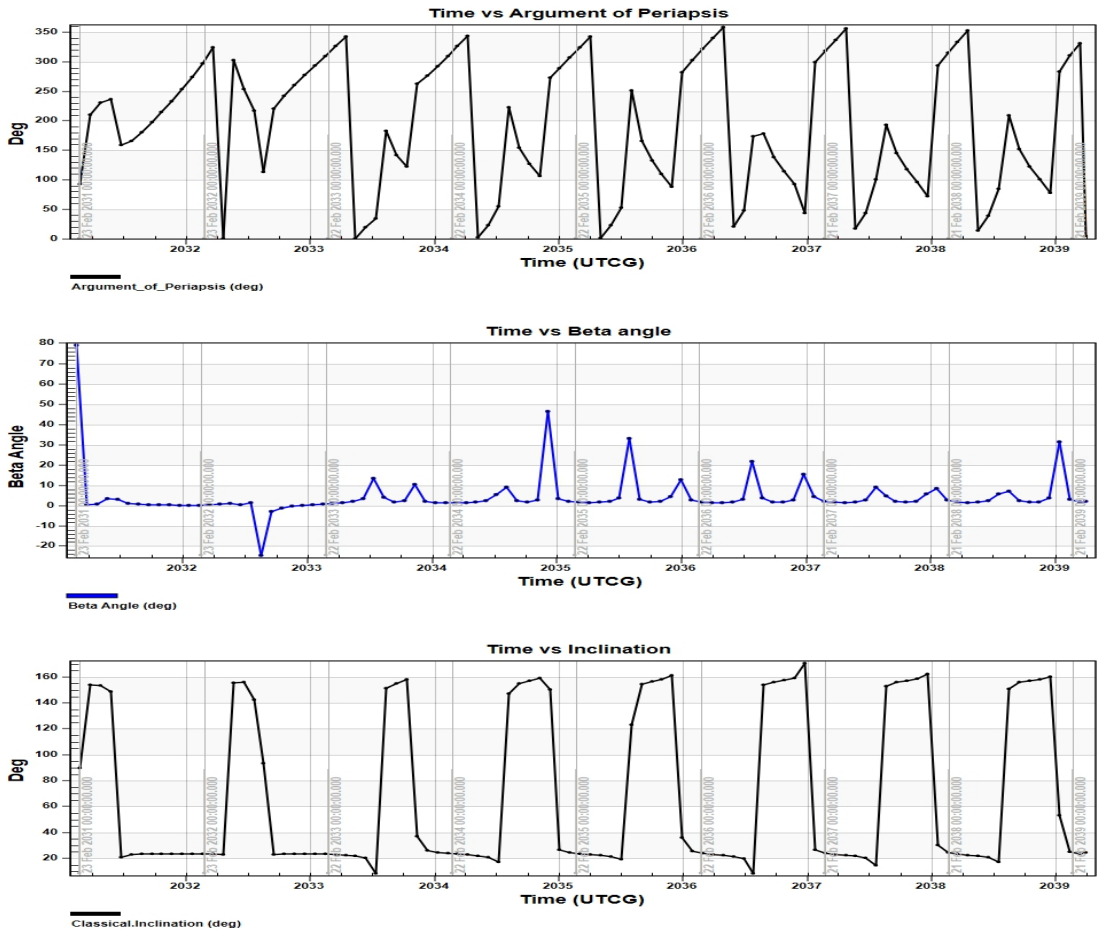

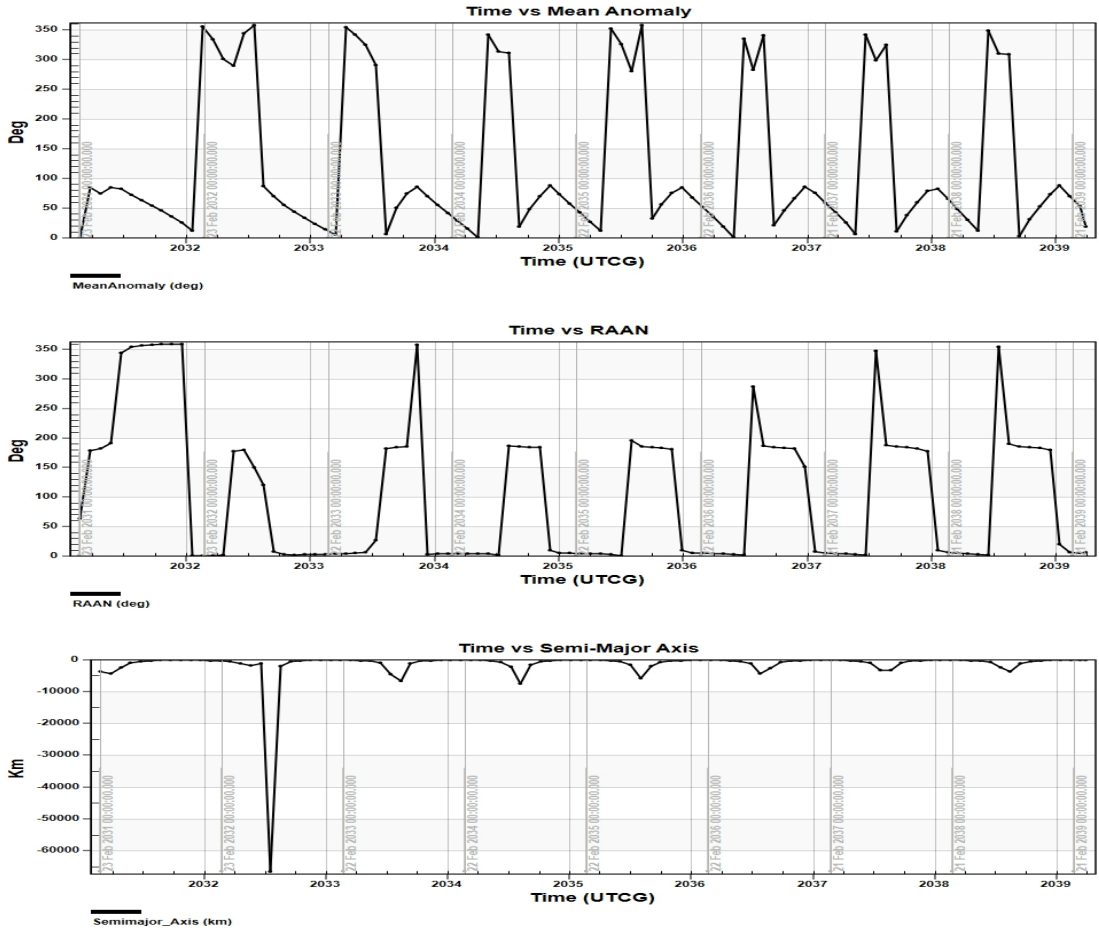

**Figure 9:** Uranian's Orbital Parameters.

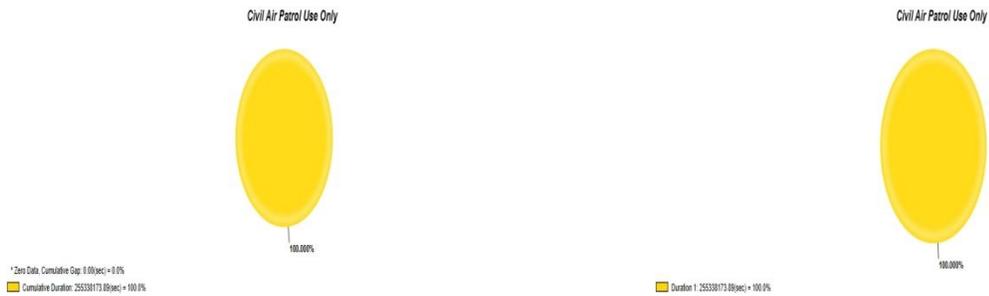

**Figure 10:** Uranian's Cumulative Sunlight Intervals.

From Jupiter to Uranus and Neptune, the optimization of the interplanetary trajectory by gravity assists makes it possible to use the gravity of Jupiter to redirect Nepta and Uranian towards Neptune and Uranus. The important features of trajectory optimization in this scientific simulation are single pulse, interplanetary destination, overflight trajectory modeling, planetary ephemeris model for the NUIP mission, which can be used to solve a variety of numerical optimization problems using the combination of methods such as collocation and implicit integration on the refinement of the adaptive programming mesh on the mathematical techniques and numerical methods used in the optimization combination to define the control parameters of the algorithm and call the sub-optimal control transcription program. The aim is also to define the structure of the problem and perform the initialization related to scaling, lower and upper limits, initial conditions, to calculate the differential equations on the right evaluate the point and path constraints display the optimal results of the solution, and create an output file using DNN models. From Earth to Jupiter, the impulsive hyperbolic injection of a circular orbit and the algorithm implemented in it assists Earth's departure path for the NUIP mission and is usually defined by a targeting specification which consists of in two times the specific orbital energy per unit mass, and the RAAN of the outgoing asymptote; provided by a spacecraft or determined with more sophisticated DNN conical trajectory analysis program that solves the Lambert problem for an interplanetary opportunity mission. This information can be used as initial assumptions for further trajectory simulations, and the hyperbolic and orbital features are in the same earth-centered inertial coordinate system. Targeting specifications are often provided or calculated in an average earth equator, and state vectors and orbital elements calculated by DNN algorithms will also be according to the J2000 coordinate system.

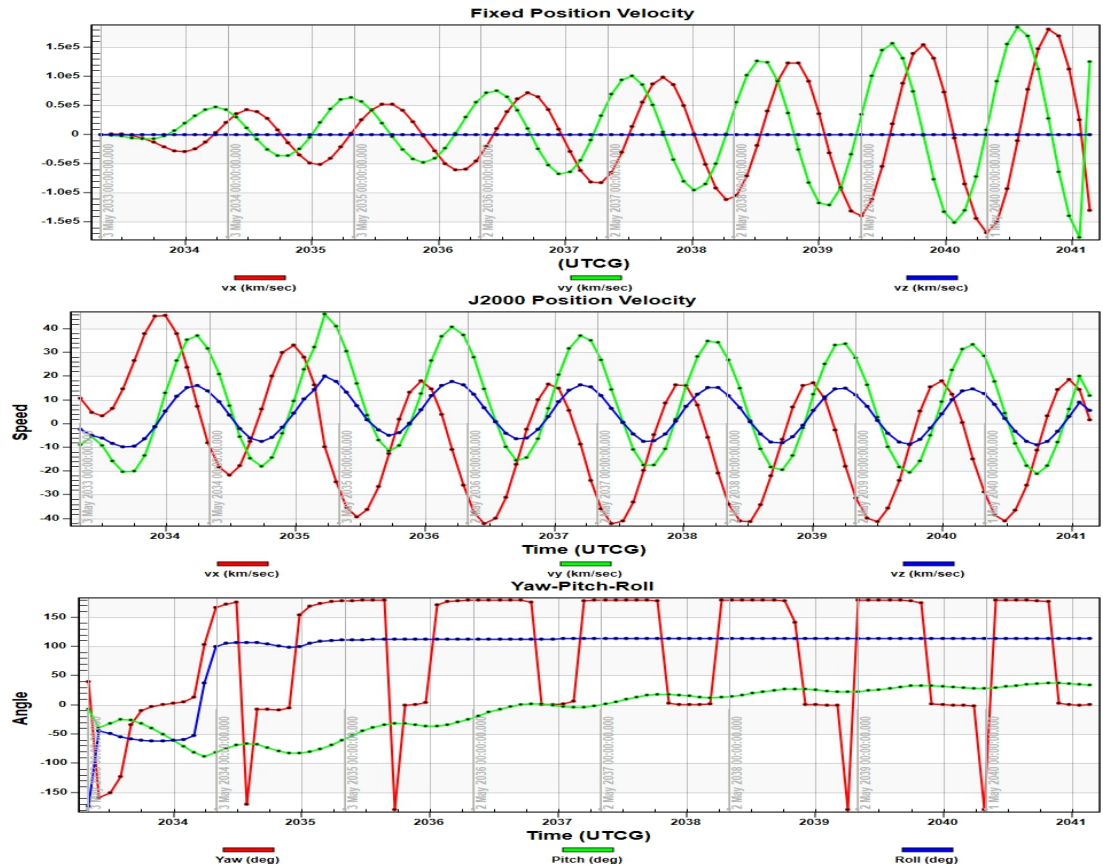

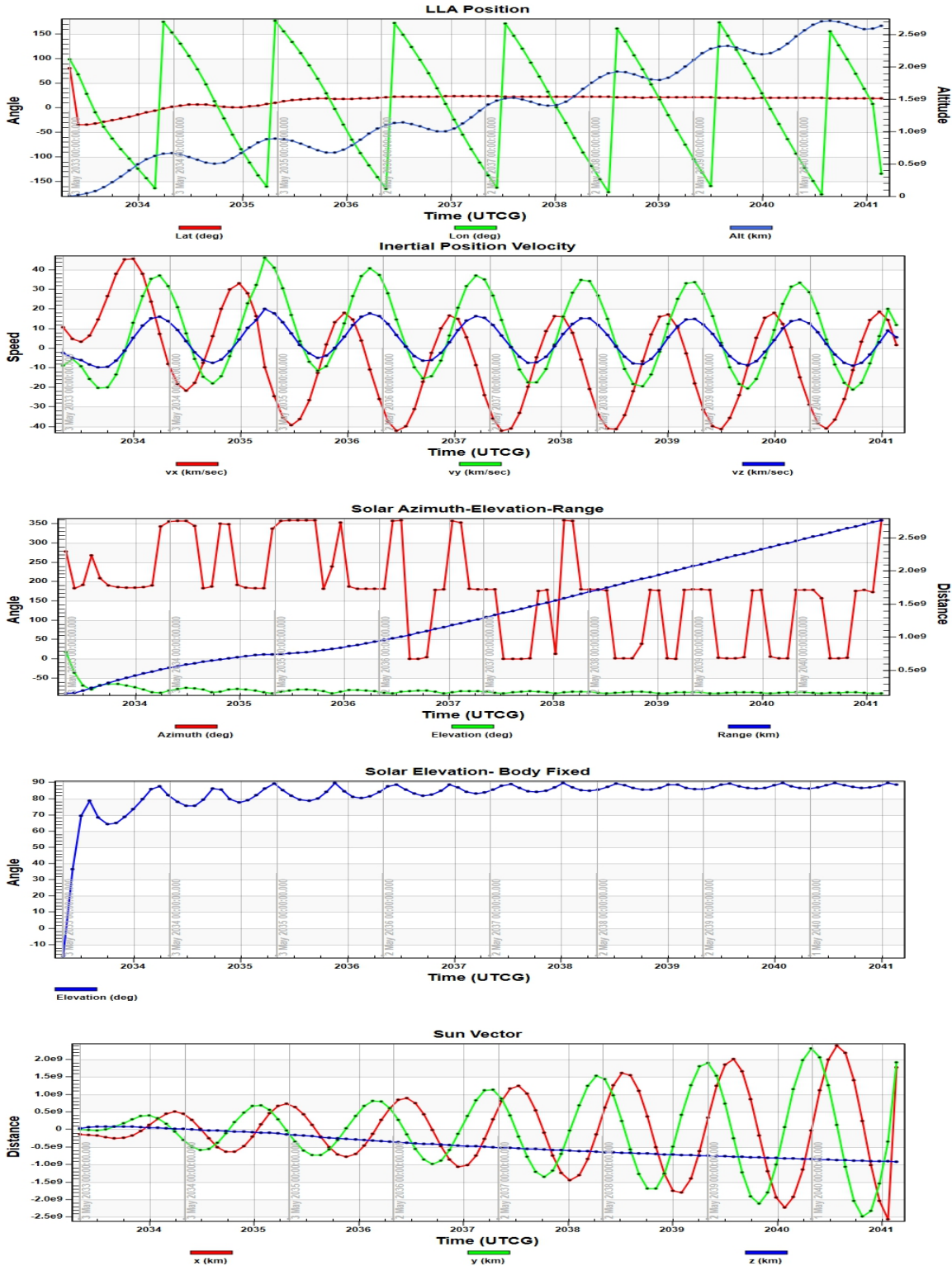

Figure 11: Nepta's Targeter Control Variables.

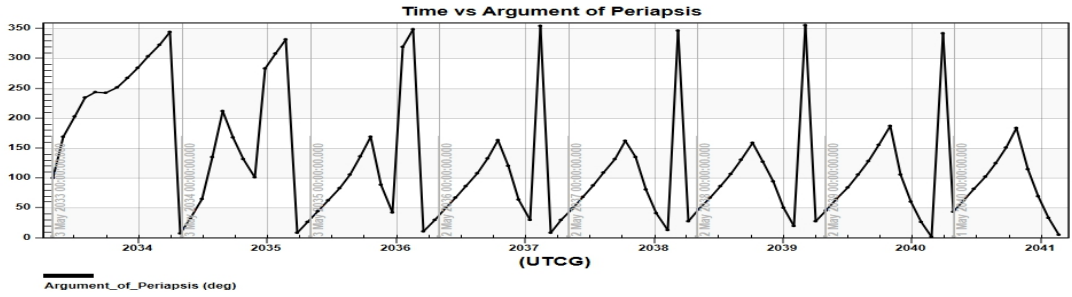
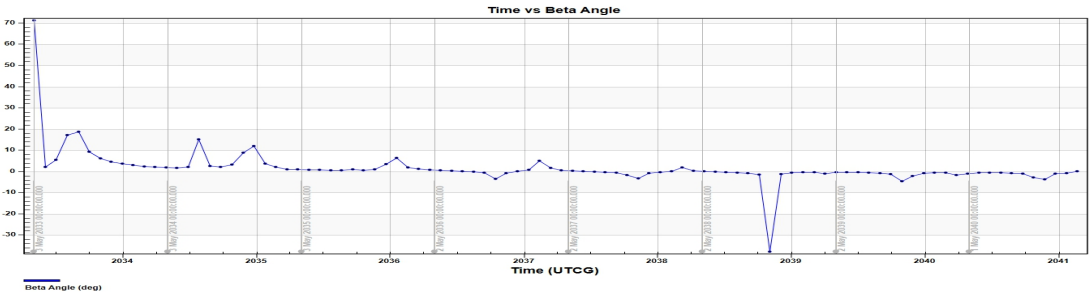
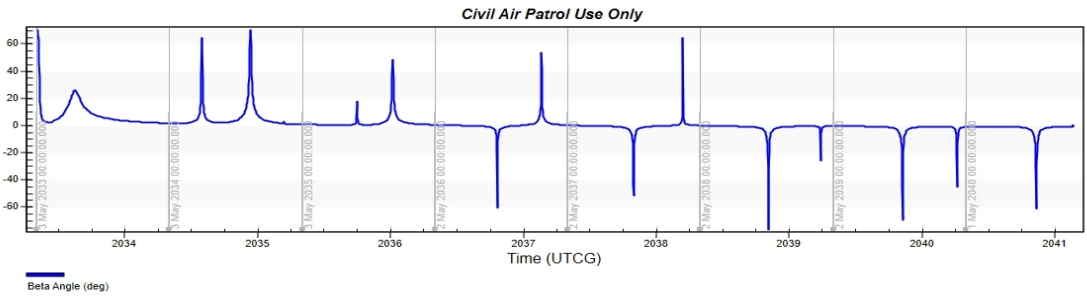
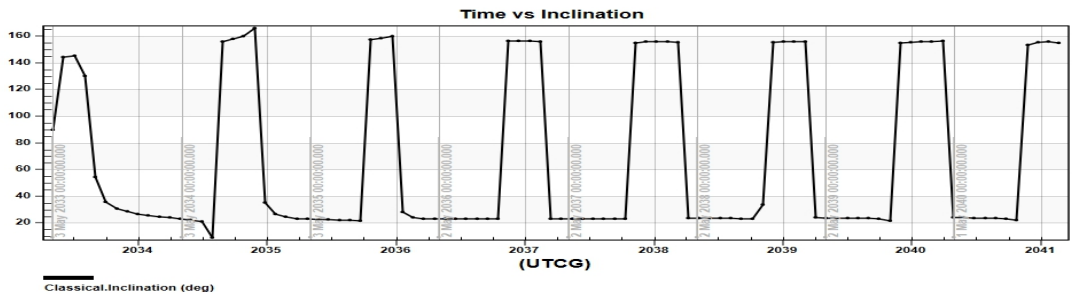
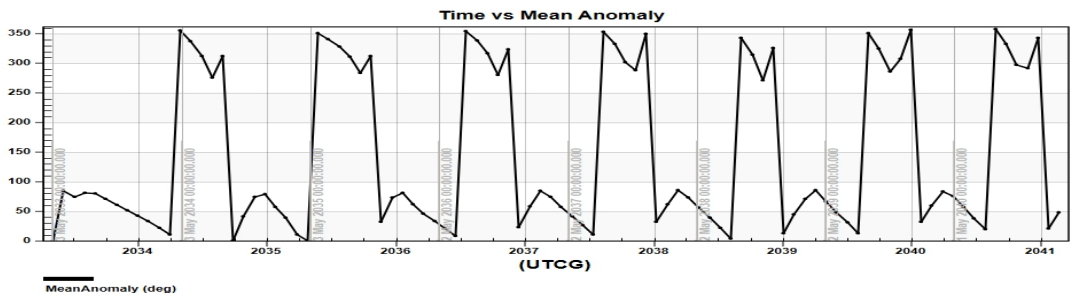

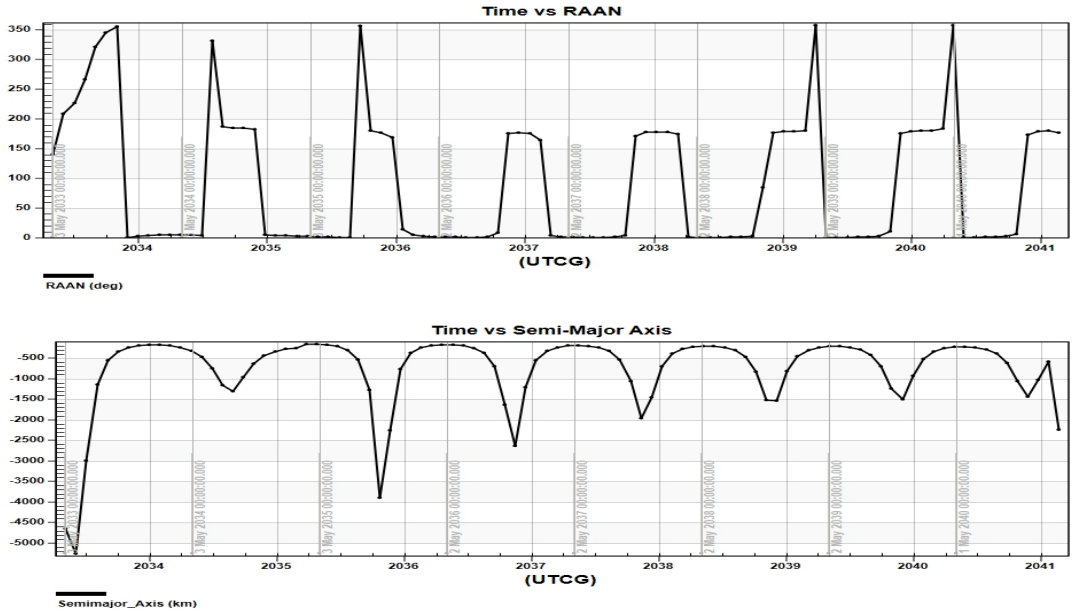

**Figure 12:** Nepta's Orbital Parameters.

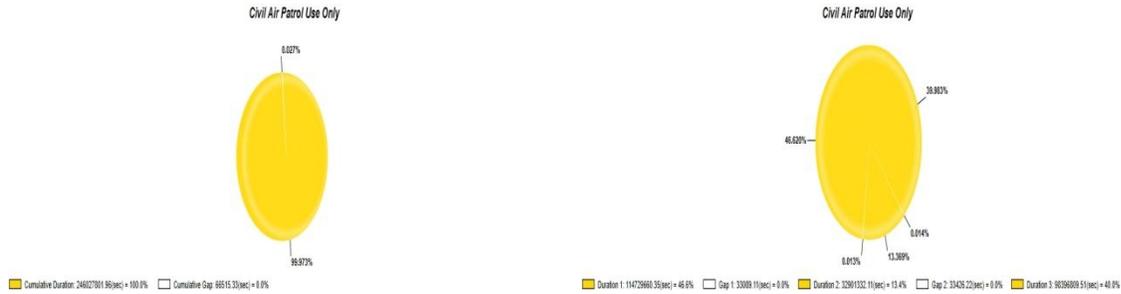

**Figure 13:** Nepta's Cumulative Sunlight Intervals.

### 5. CONCLUSIONS

In conclusion, Artificial Neural Networks can learn to recognize the input and provide the required output. CNN and LSTM are used to achieve RNN and to achieve both convolutional and regressive output. DNNs are used for path optimization in optimal guidance and are implemented in remote sensing and atmospheric detection to study ice giants' scientific properties. The NUIP mission will improve the structure and characteristics of ice giants Uranus and Neptune's atmospheric composition, including stratification, location of convective and stable regions, and internal dynamics. The NUIP mission will determine the bulk composition of the planet, including the abundances and isotopes of the elements, Helium, and the heaviest rare gases. The NUIP mission will improve knowledge on the dynamics of ice giants, determine the atmospheric heat balance of ice giants, measure the three-dimensional tropospheric flow of the planet, including winds, waves, storms, and their life cycles, and deep convective activity; characterize the structures, temporal changes of the rings and obtain a complete inventory of small moons, including source bodies embedded in dust rings and moons that could sculpt and guide dense rings and determine the surface composition of rings and moons, including organics. The NUIP mission will also map the shape and surface geology of major and minor satellites, determine the density, mass distribution, internal structure of large satellites, and, if possible, small indoor and irregular satellites, determine the composition, density, structure. The source, the spatial and temporal variability, and the dynamics of the atmospheric compositions study the solar wind-magnetosphere-ionosphere interactions and constrain the transport of plasma in the magnetosphere.

Laser-based optical communication can improve real-time data transmission at a distance of up to more than 3 AU from Earth, directing the very narrow laser beam carrying downlink optical data to Earth's tracking station. For distances greater than 3AU, it depends on a beacon signal emitted from Earth with a powerful laser. However, twice the power of the beacon is needed for each of the distances from the spacecraft. This presents a danger to the environment, and at any time, the power absorbed in the atmosphere distorts the waveform so that it is no longer even practical. Alternative methods include tracking the Earth in the infrared or referencing nearby stars with star trackers. The high data rates achievable with optical communications using modest amounts of power are largely due to direct energy in a very narrow optical beam and the use of pulse position modulation and encoding of information by the optical beam's wavelength. The scientific payload is expected to include a host of advanced sensors and antenna modules to achieve optimal telecommunications while performing flyby, encouraged by advancements in parallel computing technologies, the availability of massive tagged data, and the breakthrough in technology. During the training phase, the objective is to model the unknown functional relationship which links the given inputs to the given outputs. The inputs and labels come from a properly generated dataset, the images associated with each state are the inputs, and the optimal fuel control actions are the labels. Their performance is verified against optimal test paths never seen before by networks. This approach can be used to select actions without the need for direct filters for state estimation. Indeed, the optimal guidance is determined by processing the images, supervised machine learning algorithms are designed and tested. DNN algorithms can process precise images for classification and regression tasks, regardless of spacecraft speed, and provide a source single reasonably complete from which to search the details of optical communications with Jupiter, Uranus, and Neptune.

Table 1: Nepta's Targeter Control Variables

| Segment | Parameter | Value |
|---|---|---|
| Earth Departure | Orbit Epoch | 23 Feb 2031 |
| | C3 Energy | 105.433 km$^2$/Sec$^2$ |
| | Radius of Periapsis | 6778 km |
| | RA of Outgoing Asymptote | 243.357 deg |
| | Declination of Outgoing Asymptote | -23.5276 deg |
| | Velocity Azimuth at Periapsis | 0 deg |
| | True Anomaly | 0 deg |
| Jupiter Gravity Assist | Orbit Epoch | 10 Jul 2032 |
| | C3 Energy | 164.93 km$^2$/Sec$^2$ |
| | Radius of Periapsis | 174949 km |
| Jupiter Bplane | BDotT | 545141.91 km |
| | BDotR | -46753.75 km |
| Neptune Bplane | BDotT | 59995.0044 km |
| | BDotR | 0.75100 km |
| Neptune Arrival | Orbit Epoch | 27 Mar 2039 |
| | C3 Energy | 372.01 km$^2$/Sec$^2$ |
| | Radius of Periapsis | 44374.19 km |
| Neptune Orbit Insertion (Impulsive Maneuver) | Epoch | 27 Mar 2039 23.59.59 UTCG |
| | Delta-V Magnitude | 13854.78 m/s |
| | Finite Burn Duration(Estimated) | 5830.97 Sec |

Table 2: Nepta's Medium Fidelity Parameters

| Event | Date | Parameters |
|---|---|---|
| Launch | Feb 23 2031 | 105.433 km$^2$/Sec$^2$ (C3) |
| Jupiter Gravity Assist | Jul 10 2032 | 174949 km (Radius of Periapsis), 12.84 km/s |
| Neptune System Arrival(NOI) | Mar 27 2039 | 44374.19 km (Radius of Periapsis), 19.28 km/s |

Table 3: Nepta's Orbital Parameters (Orbit Achieved)

| Element | Values |
|---|---|
| Semi-Major Axis | 44386.67 km |
| Eccentricity | 0.0005507 |
| Inclination | 26.75 deg |
| RAAN | 219.3 deg |
| Argument of Periapsis | 131.24 deg |
| True Anomaly | 300.68 deg |
| Orbital Velocity | 12.405 km/sec |
| Orbit Period | 374.57 Min |
| Altitude | 19688.55 km |

Table 4: Uranian's Targeter Control Variables

| Segment | Parameter | Value |
|---|---|---|
| Earth Departure | Orbit Epoch | 3 May 2033 |
| | C3 Energy | 85.8259 km$^2$/Sec$^2$ |
| | Radius of Periapsis | 6778 km |
| | RA of Outgoing Asymptote | 320.56 deg |
| | Declination of Outgoing Asymptote | -33.63 deg |
| | Velocity Azimuth at Periapsis | 0 deg |
| | True Anomaly | 0 deg |
| Jupiter Gravity Assist | Orbit Epoch | 13 March 2035 |
| | C3 Energy | 51.55 km²/Sec² |
| | Radius of Periapsis | 141166.181 km |
| Jupiter Bplane | BDotT | 833416.13 km |
| | BDotR | 138501.99 km |
| Uranus Bplane | BDotT | 69999.13 km |
| | BDotR | 0 km |
| Uranus Arrival | Orbit Epoch | 17 Feb 2041 |
| | C3 Energy | 154.01 km²/Sec² |
| | Radius of Periapsis | 41848.24 km |
| Uranus Orbit Insertion (Impulsive Maneuver) | Epoch | 17 Feb 2041 23.59.59 UTCG |
| | Delta-V Magnitude | 8996.73 km/sec |
| | Finite Burn Duration(Estimated) | 5607.56 sec |

Table 5: Uranian's Medium Fidelity Parameters

| Event | Date | Parameters |
|---|---|---|
| Launch | May 03 2033 | 85.8259 km²/Sec² (C3) |
| Jupiter Gravity Assist | March 13 2035 | 141166.18(Radius of Periapsis), 7.179 km/s |
| Uranus System Arrival(UOI) | Feb 17 2041 | 41848.24(Radius of Periapsis), 12.41 km/s |

Table 6: Uranian's Orbital Parameters (Orbit Achieved)

| Element | Values |
|---|---|
| Semi-Major Axis | 41848.24 km |
| Eccentricity | 0.0000010817 |
| Inclination | 82.358 deg |
| RAAN | 172.139 deg |
| Argument of Periapsis | 331.215 deg |
| True Anomaly | 355.76 deg |
| Orbital Velocity | 11.76 km/sec |
| Orbit Period | 372.44 Min |
| Altitude | 16462.13 km |

### ABBREVIATION

1. DNN: Deep Neural Networks

2. CNN: Convolutional Neural Networks

3. RNN: Recurrent Neural Networks

4. LSTM: Long Short-Term Memory

5. NUIP: Nepta-Uranian Interplanetary

6. SIANN: Space Invariant Artificial Neural Networks

7. ANN: Artificial Neural Networks